\documentclass[conference]{IEEEtran}
\usepackage{amsfonts}
\usepackage{times}
\usepackage{graphicx}
\usepackage{latexsym}
\usepackage{dsfont}
\usepackage{amssymb}
\usepackage{amsmath}
\usepackage{cite}
\usepackage{verbatim}

\newtheorem{algorithm}{Algorithm}

\newcommand{\figref}[1]{{Fig.}~\ref{#1}}

\def\bb0{{\mathbb{0}}}

\def\ba{{\mathbf{a}}}
\def\bb{{\mathbf{b}}}

\def\bg{{\mathbf{g}}}

\def\bm{{\mathbf{m}}}

\def\bp{{\mathbf{p}}}

\def\br{{\mathbf{r}}}

\def\bv{{\mathbf{v}}}

\def\bz{{\mathbf{z}}}
\def\b0{{\mathbf{0}}}

\def\bA{{\mathbf{A}}}
\def\bB{{\mathbf{B}}}

\def\bD{{\mathbf{D}}}

\def\bR{{\mathbf{R}}}

\def\bZ{{\mathbf{Z}}}

\def\bbC{{\mathbb{C}}}

\def\bbR{{\mathbb{R}}}

\def\cA{\mathcal{A}}

\def\cC{\mathcal{C}}

\def\cE{\mathcal{E}}
\def\cF{\mathcal{F}}
\def\cG{\mathcal{G}}

\def\cM{\mathcal{M}}
\def\cN{\mathcal{N}}
\def\cO{\mathcal{O}}

\def\cS{\mathcal{S}}

\def\sfG{\mathsf{G}}

\def\sfL{\mathsf{L}}

\def\sfd{{\mathsf{d}}}

\def\sfh{{\mathsf{h}}}

\def\sf0{{\mathsf{0}}}

\def\kron{\otimes}

\usepackage[hidelinks]{hyperref}
\usepackage{cleveref}
\usepackage{physics}
\usepackage{graphicx}

\usepackage{epstopdf}
\usepackage{enumerate}
\usepackage{algorithm}
\usepackage{amsmath}
\usepackage{amssymb,amsfonts}
\usepackage[noend]{algpseudocode}
\usepackage{float}
\usepackage{color}
\usepackage{makeidx}
\usepackage{bbm}
\usepackage{lipsum}
\usepackage{balance}
\usepackage{pdf14}
\usepackage{url}
\usepackage{graphicx}

\usepackage{cleveref}
\usepackage{steinmetz}
\usepackage{varwidth}
\usepackage{textcomp}
\usepackage[usenames,dvipsnames]{xcolor}

\usepackage{caption}
\usepackage{subcaption}
\usepackage[inline]{enumitem}
\usepackage{comment}

\let\oldFootnote\footnote
\newcommand\nextToken\relax
\renewcommand\footnote[1]{%
	\oldFootnote{#1}\futurelet\nextToken\isFootnote}
\newcommand\isFootnote{%
	\ifx\footnote\nextToken\textsuperscript{,}\fi}

\makeatletter
\@addtoreset{subfigure}{row}
\makeatother

\newcommand{\sref}[1]{{Section}~\ref{#1}}

\DeclareMathOperator*{\argmax}{arg\,max}

\DeclareMathOperator\Arg{\mathrm{arg}}
\DeclareMathOperator\RE{\mathrm{Re}}
\DeclareMathOperator\IM{\mathrm{Im}}

\newcommand*{\J}{\jmath}
\newcommand*{\E}{\mathrm{e}}

\def \bEpsi{\boldsymbol{\Psi}}

\def \rm {\mathrm}
\def \bpsi{\boldsymbol{\psi}}
\newcommand{\subto}{\operatorname{s.t.}}

\usepackage{booktabs}
\usepackage{siunitx}
\usepackage{empheq}

\algnewcommand{\Initialize}[1]{%
	\State \textbf{Initialization:} \parbox[t]{.8\linewidth}{\raggedright #1}}

\begin{document}

\title{Reconfigurable Intelligent Surface Aided \\ Wireless Sensing for Scene Depth Estimation}

\author{\IEEEauthorblockN{Abdelrahman Taha, Hao Luo, and Ahmed Alkhateeb \\}
	\IEEEauthorblockA{\textit{Arizona State University}, Tempe, AZ, USA. \\ Emails: $\{$a.taha, h.luo, alkhateeb$\}$@asu.edu}
}
\maketitle

\begin{abstract}
	Current scene depth estimation approaches mainly rely on optical sensing, which carries privacy concerns and suffers from estimation ambiguity for distant, shiny, and transparent surfaces/objects. Reconfigurable intelligent surfaces (RISs) provide a path for employing a massive number of antennas using low-cost and energy-efficient architectures. This has the potential for realizing RIS-aided wireless sensing with high spatial resolution. In this paper, we propose to employ RIS-aided wireless sensing systems for scene depth estimation. We develop a comprehensive framework for building accurate depth maps using RIS-aided mmWave sensing systems. In this framework, we propose a new RIS interaction codebook capable of creating a sensing grid of reflected beams that meets the desirable characteristics of efficient scene depth map construction. Using the designed codebook, the received signals are processed to build high-resolution depth maps. Simulation results compare the proposed solution against RGB-based approaches and highlight the promise of adopting RIS-aided mmWave sensing in scene depth perception.
\end{abstract}

\section{Introduction} \label{sec:Intro}

Because of their promising coverage and spectral efficiency gains \cite{Taha2021b}, the use of reconfigurable intelligent surfaces (RISs) is envisioned as a key enabler for next-generation communication systems. 
These surfaces comprise massive numbers of nearly passive elements that interact with the incident signals in a smart way to improve the performance of such systems. 
RISs have recently started gaining interest in improving some of the wireless sensing systems \cite{Buzzi2022, li2019, hu2022}, with no application yet in scene depth estimation.
Current scene depth estimation approaches reply on optical sensing. 
While optical sensors can provide good accuracy, they suffer from some critical limitations. 
These limitations stem from the fundamental properties of the way light propagates and interacts with the elements of an environment. The accuracy of optical sensors normally degrades in scenarios of unfavorable light conditions, in the presence of shiny, dark, or transparent objects/surfaces, and in the presence of non-line-of-sight (NLoS) objects/surfaces.
Optical sensors suffer from key privacy concerns and range estimation ambiguity for distant objects/surfaces. 

To overcome these limitations, mmWave wireless sensing is a promising technology for complementing optical sensors in accurately sensing the environment.
mmWave signal propagation is not affected by interference from light sources, which can aid in recognizing shiny, dark, or transparent objects/surfaces. 
Wireless sensing systems also have fewer privacy concerns and can be well integrated with wireless communication systems \cite{Demirhan2022}. 
In \cite{Taha2021}, a mmWave MIMO based sensing framework is developed for estimating scene depth maps, under the constraints of a mmWave communication system. 
Scaling mmWave MIMO antenna arrays, however, is associated with large computational/hardware complexity and energy consumption. This limitation poses a critical challenge in scaling the spatial resolution, which motivates leveraging RISs to assist mmWave wireless sensing systems.

RIS-aided sensing systems is gaining interest in the literature.
In \cite{Buzzi2022}, a general signal model for RIS-aided target detection is studied by considering monostatic, bistatic, LOS, and NLOS scenarios.
In \cite{li2019}, the RIS-aided microwave imaging systems are proposed, where the image of the targets can be reconstructed from the receive signals.
In \cite{hu2022}, the RIS-aided RF sensing system for semantic segmentation is proposed. The semantic recognition is conducted based on the point cloud of the objects, which is reconstructed from the receive signals.
To the best of our knowledge, RIS-aided sensing systems have not yet been investigated for scene depth estimation. Accurate scene depth perception can enable some key emerging applications, including augmented and virtual reality (AR/VR) and automotive vehicles among others.

In this paper, we investigate the RIS aided wireless sensing based scene depth estimation problem. The contributions of this paper can be summarized as follows.
\begin{itemize}
	\item \textit{RIS sensing based scene depth estimation framework:} We formulate the RIS wireless sensing based scene depth estimation problem and propose a framework for building scene depth maps using RIS aided sensing systems.
	
	\item \textit{Depth map suitable RIS sensing codebook:} We propose a novel RIS interaction codebook design capable of creating a sensing grid of reflected beams that meets the desirable characteristics of efficient scene depth map construction. Given the designed codebook, the received signals are processed to build high-resolution depth maps.
\end{itemize}
Based on accurate 3D ray-tracing Wireless InSite \cite{Remcom} channels and ground truth Blender \cite{Blender} depth maps, the simulation results show the promise of adopting RIS aided mmWave sensing for scene depth estimation.

\textbf{Notation}: $\bA$ is a matrix, $\ba$ is a vector, $a$ is a scalar. $\cA$ and $\boldsymbol{\cA}$ are sets of scalars and vectors. $\|\ba \|_p$ is the p-norm of $\ba$. $\bA^T$ and $\bA^{\ast}$ are the transpose and conjugate of $\bA$. $[\bA]_{r,c}$ is the element in the $r^\rm{th}$ row and $c^\rm{th}$ column of the matrix $\bA$. $\rm{diag}(\ba)$ is a diagonal matrix with the entries of $\ba$ on its diagonal. 
$\bA \kron \bB$ is the Kronecker product of $\bA$ and $\bB$ and $\bA \odot \bB$ is their Hadamard product. $\cN(\bm,\bR)$ is a complex Gaussian random vector with mean $\bm$ and covariance $\bR$. $|\boldsymbol{\mathcal{A}}|$ is the cardinality of the set $\boldsymbol{\mathcal{A}}$. $\RE(z)$, $\IM(z)$, and $\Arg(z)$ are the real part, the imaginary part, and the phase angle of the complex number $z$. $f(t) \ast g(t)$ is the continuous-time convolution of two signals $f(t)$ and $g(t)$. $\rm{FFT}_m(\cdot)$ is the 1D FFT operation on the input matrix along its column dimension of index $m$.

\section{System and Channel Models} \label{sec:SysCh Model}

In this section, we present the adopted system and channel models for RIS aided wireless sensing systems.

\begin{figure}[t] \centerline{\includegraphics[width=0.98\columnwidth]{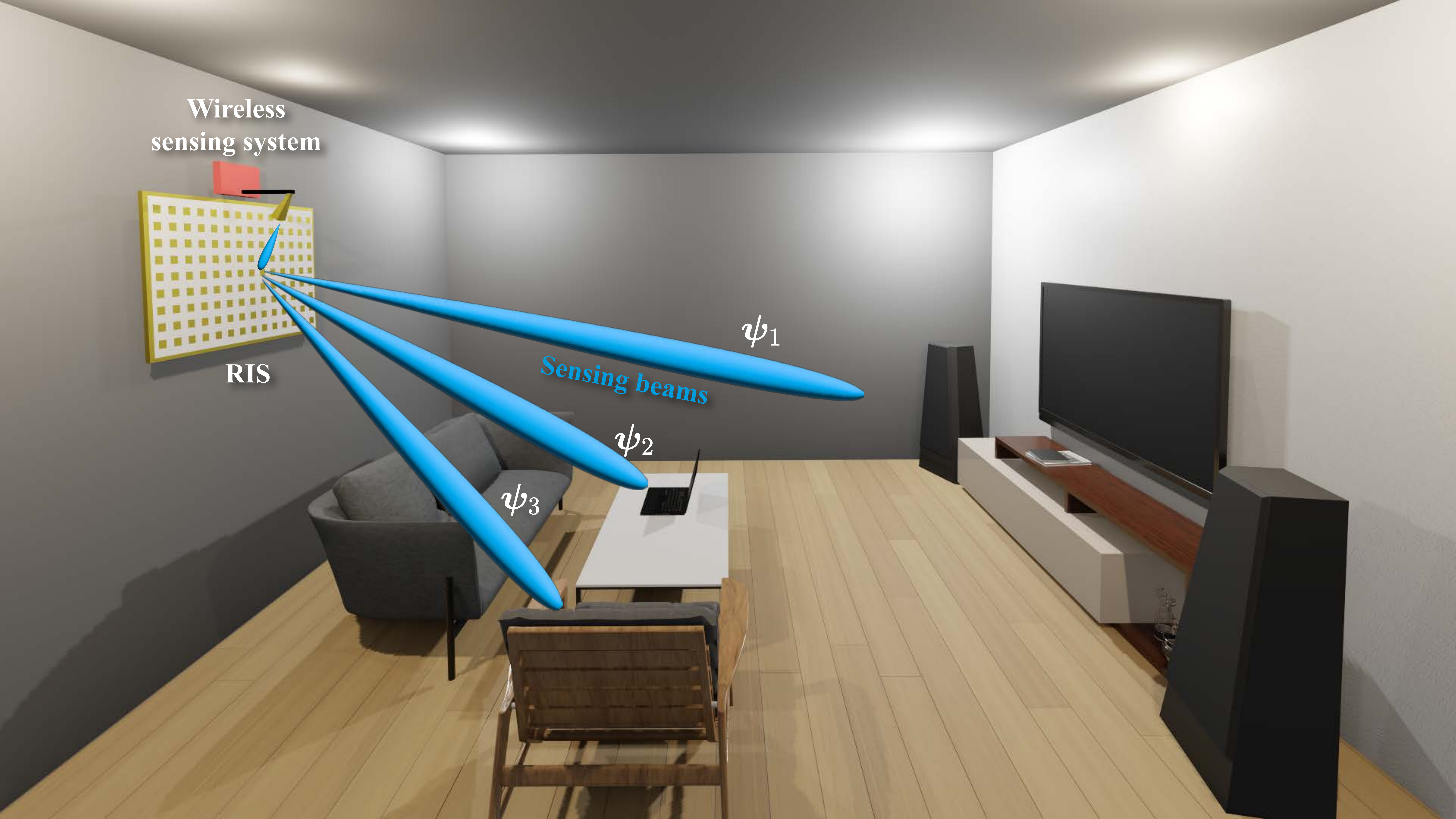}}
	\caption{The RIS aided wireless sensing system is shown. The sensing signals are transmitted to the RIS through a feeding antenna. The RIS then reflects the incident signals to the environment. The backscattered/reflected signals are then reflected by the RIS back to the sensing system, using a sensing codebook, for depth perception.} 
	\label{fig:Sys_Model}
\end{figure}

\subsection{System Model} \label{sec:Sys Model}

In this paper, we adopt a reconfigurable intelligent surface (RIS) aided mmWave wireless sensing system, as shown in \figref{fig:Sys_Model}.
The sensing system consists of a transmitter and a receiver; both are connected through a self-isolation circuitry \cite{Taha2021} to a shared single antenna, for ease of exposition. This single antenna acts as a feeding antenna illuminating the RIS for sensing purposes. The proposed solution can be extended to multi-antenna sensing transceivers.
The RIS is equipped with $N$ reconfigurable elements, where each element can be modeled as a phase shifter.
Denote the RIS interaction matrix by $\bEpsi = \rm{diag}\left({\bpsi}\right) \in\bbC^{N \times N}$, where $\bpsi=\left[\E^{\J\phi_{1}}, \dotsc, \E^{\J\phi_{N}}\right]^T$ is the interaction vector with unit modulus entries.

The sensing process proceeds as follows: (a) the sensing system transmits sensing signals to the RIS; (b) the RIS reflects these signals towards the surrounding environment, which contains $G_{\rm{tar}}$ targets; (c) the signals are reflected back to the surface by the targets; (d) the RIS reflects back these incident signals to the sensing system; (e) the sensing system processes the receive signals to achieve a sensing objective.
In this paper, our sensing objective is to estimate the depth map of the environment. For that objective, we make the following assumptions: (a) The RIS elements are not mutually correlated; (b) the channel between the sensing system and the RIS is in the near field region whereas the channel between the RIS and the targets is in the far field region; (c) the channel between the sensing system and the targets is neglected, assuming the feeding antenna radiation pattern is directional towards the RIS; (d) the RIS interaction is reciprocal when interchanging the incident signal directions with the reflected signal directions. Next, we describe the transmit and receive signal models and channel model.

\noindent \textbf{Transmit Signal Model:}
%
The adopted sensing system is a wideband FMCW radar transceiver, with a complex-baseband architecture, as detailed in \cite{ramasubramanian2017}. 
Let the radar transmit signal be a radar frame, which consists of a sequence of $M_\rm{chirp}$ repeated chirp signals with a chirp repetition interval of $T_{\rm{PRI}}$ seconds.
Let $a_\rm{BP}(t)  \in \bbR$ be the bandpass transmit signal of a single chirp, with a duration of $T_{\rm{active}}$ seconds, a bandwidth
of $\rm{BW}= S T_{\rm{active}}$, a chirp slope of $S$, and a starting chirp frequency of $f_0$.
The signal $a_\rm{BP}(t)$ can then be formulated as
\begin{equation}
	a_\rm{BP}(t) =
	\begin{cases}
		\cos\left(2\pi f_{0}t + \pi S t^2\right) & 0\leq t \leq T_{\rm{active}},\\
		0 & \text{otherwise}.
	\end{cases}
\end{equation}
The transmit signal of a radar frame $x_\rm{BP}(t)$ can be defined as
\begin{align}
	x_\rm{BP}(t) &=\sqrt{\cE_\rm{T}} \, \sum_{c=0}^{M_\rm{chirp}-1} a_\rm{BP}(t-cT_{\rm{PRI}})  \\
	& = \RE\left( x(t) \, \E^{\J 2\pi f_0 t } \right), t\in\bbR_{\geq 0},
\end{align}
where $\cE_\rm{T}$ is the transmit signal energy and $x(t) \in \bbC$ is the complex-valued lowpass-equivalent transmit signal.

\noindent  \textbf{Receive Signal Model:}
%
For the channel model, we adopt the extended Saleh-Valenzuela wideband geometric channel model \cite{Taha2021}. After traveling through the bandpass channel, the receive bandpass signal $y_\rm{BP}(t) = \RE(y(t) \E^{\J 2\pi f_0 t })$ can be modeled in terms of its lowpass-equivalent $y(t) \in \bbC$, which can be defined as 
\begin{align}
	y(t) &= x(t) \ast h(t) + w(t) \\ 
	&= \sum_{g=1}^{G_\rm{tar}} \sum_{\ell=1}^{L_g} \sfh_{g,\ell} x(t - \xi_{g,\ell} ) + w(t),
\end{align}
where $h(t)$ is the lowpass-equivalent channel and $w(t) \sim \cN(0,\sigma_{w}^2)  \in \bbC$ is the receive noise with variance $\sigma_{w}^2$. $L_g$ is the number of channel paths interacting with the $g^\rm{th}$ target. $\sfh_{g,\ell} \in \bbC$ is the complex channel gain of the $\ell^\rm{th}$ channel path of the $g^\rm{th}$ target, which is detailed in \sref{sec:Ch Model}.
The propagation delay is $\xi_{g,\ell}= R_{g,\ell}/\varsigma$, where $\varsigma$ is the speed of light. $R_{g,\ell}$ denotes the total propagation distance traveled by the $\ell^\rm{th}$ channel path of the $g^\rm{th}$ target (with one or multiple interactions with the environment).

To construct the receive baseband intermediate frequency (IF) signal \cite{ramasubramanian2017}, the receive signal $y_\rm{BP}(t)$ is first mixed with two versions of the transmit signal $x_\rm{BP}(t)$, one with a $-90^\circ$ phase shift difference. Then, the outputs of the mixers pass through low-pass filters and ADCs to generate the in-phase signal $I[s,c]$ and the quadrature-phase signal $Q[s,c]$, for the ADC sample $s \in \cS, \cS=\left\{0,1,\ldots,\left(M_\rm{sample}-1\right)\right\}$, and for the chirp $c \in \cC, \cC=\left\{0,1,\ldots,\left(M_\rm{chirp}-1\right)\right\}$. $M_\rm{sample}$ is the number of ADC samples per chirp.
Let $b[s,c]$ denotes the discrete-time equivalent of a continuous-time signal $b(t)$, sampled at time $t=sT_{\rm{S}}+cT_{\rm{PRI}}$,
$T_{\rm{S}}=1/F_{\rm{S}}$, where $F_{\rm{S}}$ is the ADC sampling frequency. 
The receive baseband IF digital signal, $z[s,c] = I[s,c] + \J \, Q[s,c]$, can be formulated as 
\begin{multline}
	z[s,c] = \sum_{g=1}^{\sfG_{\rm{tar}}} \sum_{\ell=1}^{L} \sqrt{\rho_{g,\ell}} \, \E^{- \J \vartheta_{g,\ell}} \, \E^{+ \J \, \Xi_{g,\ell}} +  w[s,c] \E^{\J \chi[s]},
\end{multline}
where $\chi[s] = 2\pi f_{0}t_{\rm{fast}} + \pi S t_{\rm{fast}}^2 $ and $t_{\rm{fast}}=sT_{\rm{S}}$.
The channel path receive power and phase are $\rho_{g,\ell}= \cE_\rm{T} |\sfh_{g,\ell}|^{2}$ and $\vartheta_{g,\ell}=\Arg\left( \sfh_{g,\ell}\right)$, respectively. 
The proof is left for a journal extension due to space limitations.
The phase term $\Xi_{g,\ell}$ contains range information of the targets, which is defined as
\begin{equation}
	\Xi_{g,\ell}  =  2\pi  \left( f_0 \xi_{g,\ell}  +  S t_{\rm{fast}} \xi_{g,\ell} -  \tfrac{S}{2} \xi_{g,\ell}^{2} \right).
\end{equation}
Next, we describe the complex channel gain model $\sfh_{g,\ell}$.

\subsection{Channel Model} \label{sec:Ch Model}
For RIS aided radar channel modeling, we adopt and extend on the channel model of the non-line-of-sight monostatic radar configuration detailed in \cite{Buzzi2022}. Different from the model in \cite{Buzzi2022}, we adopt a multi-path geometric channel model where each channel path can experience one or multiple interactions in the environment, which consists of multiple targets. The complex channel gain $\sfh_{g,\ell} \in \bbC$ can be modeled as \cite{Buzzi2022}
\begin{align}
	&\sfh_{g,\ell} =  \underbrace{(\bg^{T} \bEpsi \bv(\bar{\theta}_{g,\ell} ) \bar{\gamma}_{g,\ell})}_{\text{Radar} \rightarrow \text{RIS} \rightarrow \text{Target}} \times \underbrace{(\bg^{T} \bEpsi \bv(\ddot{\theta}_{g,\ell}) \ddot{\gamma}_{g,\ell})}_{\text{Target} \rightarrow \text{RIS} \rightarrow \text{Radar}}, \\
	& \!\!\! = \bar{\gamma}_{g,\ell} \left(\left( \bg \odot \bpsi \right)^{T}  \bv \left( \bar{\theta}_{g,\ell} \right) \right) \times \ddot{\gamma}_{g,\ell} \left( \left( \bg \odot \bpsi \right)^T  \bv \left( \ddot{\theta}_{g,\ell}  \right) \right)\!, \!\!
\end{align}
where  $\bg \in \bbC^{N}$ is the normalized near-field forward/backward channel vector between the radar feeding antenna and the RIS elements. The normalization is relative to the scalar channel passing through the RIS reference element, whose complex gain is included in the definitions of $\bar{\gamma}_{g,\ell}, \ddot{\gamma}_{g,\ell}$.
%
The far-field transmit/receive RIS array response vector is $\bv(\cdot)\in \bbC^{N}$.
Let an angle notation of $\varphi$ denote the set of the azimuth and zenith angles, $\varphi=\{\varphi^{\rm{az}},\varphi^{\rm{ze}}\}$.
$\bar{\theta}_{g,\ell}$ (and $\ddot{\theta}_{g,\ell}$) are the azimuth and zenith angles of departure (and arrival) of the $\ell^\rm{th}$ channel path of the $g^\rm{th}$ target, relative to the RIS reference element. 
$\cG(\varphi)$ is the transmit/receive gain of the feeding antenna in the direction $\varphi$.
$\bar{\gamma}_{g,\ell}, \ddot{\gamma}_{g,\ell} \in \bbC$ are the two-hop forward and backward complex channel path gains, including the propagation between the radar transceiver and the RIS reference element, and the propagation between the RIS reference and the $g^\rm{th}$ target.


The normalized channel vectors $\bg$ can be represented as  \cite{Buzzi2022}
\begin{equation}
	[\bg]_n= \sqrt{ \frac{\cG(\bar{\Omega}_{n})\zeta(\bar{\omega}_{n}, \bar{\theta}_{g,\ell})\delta_1^2}{\cG(\bar{\Omega_{1}})\zeta(\bar{\omega}_{1}, \bar{\theta}_{g,\ell})\delta_{n}^{2}} } \cdot \E^{-\J2\pi(\delta_{n} - \delta_1)/\lambda},
\end{equation}
where $n \in \{1,\dotsc,N\}$ and $\lambda  = \tfrac{\varsigma}{f_0}$ is the operating wavelength. $\delta_n$ is the distance between the radar feeding antenna and the $n^\rm{th}$ RIS element, where $\delta_1$ represents the distance with respect to the RIS reference element. Let the vector of distances between the radar feeding antenna and the RIS elements be $\boldsymbol{\delta} = [\delta_1,\dotsc,\delta_N]^T$.
$\bar{\Omega}_{n}$ (and $\ddot{\Omega}_{n}$) are the azimuth and zenith angles of departure (and arrival), relative to the radar feeding antenna, for the propagation between the radar transceiver and the $n^\rm{th}$ RIS element.
$\bar{\omega}_{n}$ (and $\ddot{\omega}_{n}$) are the azimuth and zenith angles of arrival (and departure), relative to the $n^\rm{th}$ RIS element, for the propagation between the radar transceiver and the $n^\rm{th}$ RIS element.  
$\zeta(\varphi_{\rm{in}}, \varphi_{\rm{out}})$ is the radar cross-section gain of an RIS element towards the direction $\varphi_{\rm{out}}$, when illuminated from the direction $\varphi_{\rm{in}}$, which is modeled in \cite{Buzzi2022}.
%
%
The two-hop forward and backward complex channel path gains are defined as \cite{Buzzi2022}
\begin{align}
	\bar{\gamma}_{g,\ell} &= \sqrt{ \frac{\cG(\bar{\Omega}_{1})\zeta(\bar{\omega}_{1}, \bar{\theta}_{g,\ell})}{(4\pi)^2 \delta_1^2 {\bar{d}_{g,\ell}^{2}} {\bar{\sfL}}_{g,\ell}}  } \,\, \E^{-\J2\pi(\delta_1 + \bar{d}_{g,\ell})/\lambda}, \label{eq:forward_gamma} \\
	\ddot{\gamma}_{g,\ell} &=  \sqrt{ \frac{\sigma_{g}\zeta(\ddot{\theta}_{g,\ell}, \ddot{\omega}_{1})\cG(\ddot{\Omega}_{1})\lambda^2}{ (4\pi)^3 {\ddot{d}_{g,\ell}^{2}} \delta_1^2 {\ddot{\sfL}}_{g,\ell}  } }    \,\, \E^{-\J2\pi(\ddot{d}_{g,\ell} + \delta_1)/\lambda}, \label{eq:backward_gamma}
\end{align}
where $\bar{d}_{g,\ell}, \ddot{d}_{g,\ell}$ are the forward and backward traveling distance of the $\ell^\rm{th}$ path, between the RIS reference element and the $g^\rm{th}$ target, which can be related to the total propagation distance such that $R_{g,\ell} = 2\delta_1 + \bar{d}_{g,\ell} +  \ddot{d}_{g,\ell}$.
$\sigma_{g}$ is the radar cross-section gain of the $g^\rm{th}$ target.
${\bar{\sfL}}_{g,\ell}, {\ddot{\sfL}}_{g,\ell}$ are forward and backward loss factors for any additional attenuation.

\section{Problem Formulation} \label{sec:Prob Def}

In this paper, our objective is to efficiently estimate the depth map of the surrounding environment using the RIS-aided wireless sensing system described in \sref{sec:SysCh Model}. 

\subsection{Problem Definition} 

Following the depth map definition in \cite{Taha2021}, the depth map, $\bD_{\rm{map}} \in \bbR^{M_\rm{h} \times M_\rm{w}}$, can be defined as an image of resolution $M_\rm{w}$ pixels wide and $M_\rm{h}$ pixels high, where the value of each pixel denotes the smallest depth between the RIS reference element and the targets/surfaces in this pixel. The total number of pixels in the depth map is $M_\rm{res}=M_\rm{w} M_\rm{h}$. 
Through (a) effectively scanning the environment using several RIS interaction vectors and (b) processing the receive signals, the RIS aided sensing system can construct the depth map. 

To scan the environment, we define a sensing codebook of RIS interaction vectors, $\boldsymbol{\cF}=\{\bpsi_m: m \in \cM, \cM = \{0,\ldots,M-1\}\}$.
Each RIS interaction vector aids in the transmission and reception of a single chirp signal, when directed towards a certain direction in the environment.
For the $m^\rm{th}$ interaction vector, $\bpsi_m$, the complex channel gain $\sfh_{g,\ell}[m]$, $s \in \cS$, $m \in \cM$, can be expressed as
\begin{multline}
	\sfh_{g,\ell}[m] = \bar{\gamma}_{g,\ell}\left(\left( \bg \odot \bpsi_m \right)^{T}  \bv \left( \bar{\theta}_{g,\ell} \right) \right) \times \\ \ddot{\gamma}_{g,\ell} \left( \left( \bg \odot \bpsi_m \right)^T  \bv \left( \ddot{\theta}_{g,\ell}  \right) \right). \label{eq:PS_channel_m}
\end{multline}
The receive IF digital signal can then be defined as
\begin{multline}
	z[s,m] = \\ \underbrace{\sum_{g=1}^{\sfG_{\rm{tar}}} \sum_{\ell=1}^{L} \sqrt{\rho_{g,\ell}[m]} \, \E^{- \J \vartheta_{g,\ell}[m]} \, \E^{+ \J \, \Xi_{g,\ell}}}_\text{Receive signal} +  \underbrace{w[s,m] \E^{\J \chi[s]}}_\text{Noise}. \label{eq:PS_receive_sensing_scalar}
\end{multline}
where $\rho_{g,\ell}[m]= \cE_\rm{T} |\sfh_{g,\ell}[m]|^{2}$ and $\vartheta_{g,\ell}[m]=\Arg\left( \sfh_{g,\ell}[m]\right)$. 
By stacking the $S$ receive ADC samples, we can construct the receive sensing vector, $\bz[m] \in \bbC^{M_\rm{sample}}$, corresponding to the transmission of a single chirp signal using one RIS interaction vector, $\bz[m] = \left[ z[0,m],  \dotsc, z[M_\rm{sample}-1,m] \right]^T$.
If $M$ radar chirps (a single radar frame) are transmitted and received via $M$ RIS interaction vectors, the aggregated receive sensing signal matrix, 
$\bZ \in \bbC^{ M_\rm{sample} \times M}$, can be expressed as
\begin{equation}
	\bZ=\left[\bz[0], \bz[1], \dotsc, \bz[M-1] \right]. \label{eq:sensing_matrix}
\end{equation}

Next, to estimate the depth map, we define a post-processing function $\bp(.)$.
Given the receive matrix $\bZ$ with the RIS sensing codebook $\boldsymbol{\cF}$, the estimated depth map can be written as
\begin{equation}
	\mathbf{\widehat{D}}_{\rm{map}}=\bp(\bZ;\boldsymbol{\cF}).
\end{equation}
Our objective is to minimize the estimation error between the estimated depth map $\mathbf{\widehat{D}}_{\rm{map}}$ and the actual depth map $\bD_{\rm{map}}$.
For this reason, we adopt the root-mean squared error (RMSE) and the mean absolute error (MAE) as the performance metrics, which are defined as \cite{Taha2021}
\begin{equation}
	\Delta_{\rm{RMSE}}= \left(  \frac{1}{M}\norm{  \bD_{\rm{map}}-\bp(\bZ;\boldsymbol{\cF})  }_{2}^{2} \right)^{1/2}  ,
\end{equation}
\begin{equation}
	\Delta_{\rm{MAE}}=\frac{1}{M}\norm{\bD_{\rm{map}}-  \bp(\bZ;\boldsymbol{\cF})   }_{1}^{2}.
\end{equation}

\subsection{Main Challenges} 
Estimating scene depth maps using mmWave sensing systems suffer from the following challenges.

\noindent \textbf{1. Codebook design:}
To build RIS-based depth maps capable of complementing RGB-D based depth maps, the RIS interaction codebook needs to be designed to reflect the incident signals in the directions of the full rectangular grid of typical depth optical sensors. Classical RIS codebooks \cite{Taha2021b}, however, are designed based on DFT codebooks which forms parabolic grids instead of rectangular grids. In addition, mmWave MIMO based sensing codebooks, as detailed in \cite{Taha2021}, can not be adopted as RIS sensing codebooks.

\noindent \textbf{2. Low-resolution depth maps:}
mmWave MIMO based depth map estimation has been investigated for wireless AR/VR systems \cite{Taha2021}. Scaling mmWave antenna arrays, however, is associated with large computational/hardware complexity and energy consumption. This limitation poses a prominent challenge in scaling the spatial resolution of the depth maps.

\noindent \textbf{3. Inter-target and inter-path interferences:}
When sensing the depth of a certain region of interest (represented by a single pixel), the best scenario is when only a single target exist in that region of interest, and that target backscatters a single-bounce path to the receiver. In practice, however, it can be hard to differentiate the receive signals from multiple targets that are close to each others. The incident signals on a each target can also experience multiple bounces in directions away from the desired direction, before reaching the receiver. The challenge is how to design the RIS aided sensing solution to detect the desired channel path while filtering out the undesired channel paths \cite{Taha2021}.
In the next section, we present our proposed solution to address these scene depth estimation challenges.

\section{Proposed Solution} \label{sec:prop_soln}
In this section, we introduce a comprehensive framework for scene depth estimation using RIS aided sensing systems.

\subsection{Key Idea}

Because of the massive number of the nearly-passive RIS elements, these surfaces can adopt fine-grained reflection beams while scanning the environment, enabling high-resolution sensing grids using energy-efficient architectures \cite{Taha2021b}.  
In addition, RIS aided sensing systems can filter out more undesired paths than the ones filtered out by mmWave MIMO based sensing systems, without leveraging any elaborate post-processing functions (as opposed to the ones used in \cite{Taha2021}). One possible reason is that an RIS interaction matrix is designed to focus the reflection in one desired direction and the reception from the same direction; any channel path arriving back to the RIS from a direction other than the desired direction is reflected away from the radar receiver.
Also, for AR/VR systems, the post-processing sensing tasks can be offloaded from the AR/VR devices to the RIS aided wireless sensing systems --- a significant advantage for AR/VR appplications. 
For these reasons, we propose an RIS aided sensing based scene depth map estimation solution capable of further improving the depth perception of the surrounding environment compared to existing RGB based depth map estimation solutions \cite{hu2019,ranftl2021}.
Next, we formulate the main elements of our proposed RIS aided sensing framework, namely the RIS sensing codebook design and the scene depth estimation.

\subsection{RIS Sensing Codebook Design} \label{ssec:CB_design}
Our objective for the RIS interaction codebook design is to construct a sensing grid of reflected directions that fits the rectangular grid of a depth camera.
%
Assume the RIS is employing a uniform planar array (UPA) structure in the $x$-$z$ plane. The RIS is then equipped with $N_\rm{H}$ elements on the $x$-axis (the horizontal axis) and $N_\rm{V}$ elements on the $z$-axis (the vertical axis), where $N=N_\rm{H} N_\rm{V}$. In such case, the far-field RIS array response vector $\bv\left( \varphi \right)$, in the direction $\varphi=\left\{\varphi^\rm{az}, \varphi^\rm{ze}\right\}$, can then be formulated as
\begin{align}
	\bv\left(  \varphi \right) = \bv_z \left( \varphi \right) \kron \bv_x \left( \varphi\right).
\end{align}
where $\bv_x(.)$ and $\bv_z(.)$ represent the elemental array response vectors in the $x$ and $z$ directions, and are expressed as 
\begin{align}
	\bv_x\left(\varphi \right) &=\left[1, \E^{\J \kappa\sfd  \cos(\varphi^\rm{az}) \sin(\varphi^\rm{ze}) }, \ldots\right. \nonumber \\ &\hspace{8pt}\quad\quad \left. \ldots, \E^{\J \kappa\sfd (N_\rm{H}-1) \cos(\varphi^\rm{az}) \sin(\varphi^\rm{ze}) } \right]^T,  \\
	\bv_z\left(\varphi \right)&=\left[1, \E^{\J \kappa\sfd \cos(\varphi^\rm{ze}) }, \ldots, \E^{\J \kappa\sfd (N_\rm{V}-1)\cos(\varphi^\rm{ze}) } \right]^T,
\end{align}
where $\kappa=\tfrac{2\pi}{\lambda}$ is the wave number and $\sfd$ is the RIS element spacing in meters. 
%
For simplicity of scene definition, let the horizontal direction of the depth map be parallel to the $x$-axis, and its vertical direction be parallel to the $z$-axis. Let the RIS reference element --- the focal point of the scene depth map --- be the origin of the rectangular coordinate system. In such case, the depth of a target is measured by the $y$-coordinate of the $x$-$z$ plane of that target.
Consider an oversampled RIS interaction codebook of $M=\overline{N}_\rm{V}\overline{N}_\rm{H}$ beams, where  $\overline{N}_\rm{V}={N}_\rm{V} F_\rm{V}^\rm{OS}$ and $\overline{N}_\rm{H}={N}_\rm{H} F_\rm{H}^\rm{OS}$. ${F}_\rm{V}^\rm{OS}, {F}_\rm{H}^\rm{OS}$ are the oversampling factors in the horizontal and vertical dimensions. 

Now, we explain how to design the RIS interaction matrix to reflect the incident signal into a certain direction. From \eqref{eq:PS_receive_sensing_scalar}, the receive signal from a target in a certain direction can become more distinguishable from the ones received from targets in other directions by controlling their respective channel gains, $|\sfh_{g,\ell}|$. More specifically, to distinguish more the receive signal gain of the $\ell^\rm{th}$ channel path of the $g^\rm{th}$ target, the RIS interaction vector $\bpsi^\star$ can be designed as
\begin{align}
	\bpsi^{\star} = \argmax_{\bpsi} & |\sfh_{g,\ell}| \\ = \argmax_{\bpsi} & \! \left| (( \bg \odot \bpsi )^{T}  \bv ( \bar{\theta}_{g,\ell} ) ) ( ( \bg \odot \bpsi )^T  \bv ( \ddot{\theta}_{g,\ell} ) ) \right|, \nonumber \\
	\subto & \left| \left[ \bpsi\right]_n  \right|=1, ~ \forall n \in \{1,\dotsc,N\}.
\end{align}
Note that we are only interested in distinguishing single-bounce paths to estimate the depth correctly \cite{Taha2021}, $\bar{\theta}_{g,\ell} = \ddot{\theta}_{g,\ell} =  \theta_{g,\ell}$. In such case, the optimization problem is reduced to
\begin{align}
	\bpsi^{\star} = \argmax_{\bpsi} \hspace{3pt} & \hspace{3pt} \left| ( \bv (\theta_{g,\ell})  \odot \bg)^{T} \bpsi \right|^2, \\ 
	\subto  \hspace{3pt} & \hspace{3pt} \left| \left[ \bpsi\right]_n  \right|=1, ~ \forall n \in \{ 1,\dotsc,N\},
\end{align}
Assume prior knowledge of (i) the distance vector $\boldsymbol{\delta}$ between the radar feeding antenna and the RIS elements and (ii) the direction specified by $\theta_{g,\ell}$. The RIS interaction vector $\bpsi^\star$ can then be designed using equal-gain conjugate beamforming as
\begin{multline}    
	\bpsi^{\star} =  ( \bv ( \theta_{g,\ell} )  \odot \Arg\left(\bg\right) )^{\ast} \\ = ( \bv ( \theta_{g,\ell} )  \odot  \E^{-\J2\pi(\boldsymbol{\delta} - \delta_1)/\lambda} )^{\ast}. \label{eq:int_vec_star}
\end{multline}


Next, we explain how to design the RIS interaction codebook to construct a sensing grid of reflected directions that fits the rectangular grid of a depth camera. More specifically, let $\cO$ be the set of spherical coordinate angles representing the grid point directions from the desired rectangular grid, such that $\cO = \{\theta_m\}_{m=0}^{M-1}$. We adopt the design of the set $\cO$ from [6, Sec. VII, Eq. 19] to eliminate any grid mismatch distortion. The set $\cO$ can be completely described using the scene field of view $\! \rm{FoV}$, the aspect ratio of the depth map $\! A_\rm{R}$, and the number of horizontal and vertical grid points $\! \overline{N}_\rm{H},\overline{N}_\rm{V}$, as detailed in \cite{Taha2021}.
For the $m^\rm{th}$ grid point pointing towards the direction $\theta_m \in \cO$, the RIS vector $\bpsi_{m}^{\star}$ can be designed as
\begin{align}
	\bpsi_{m}^{\star} = \argmax_{\bpsi_{m}} \hspace{3pt} & \hspace{3pt} |\sfh_{g,\ell}[m]| \\
	\subto  \hspace{3pt} & \hspace{3pt} \left| \left[ \bpsi_m\right]_n  \right|=1, ~ \forall n \in \{1,\dotsc,N\}, \\
	\bpsi_{m}^{\star} = \Big( \bv(\theta_m) \,\,  & \odot \,\,  \E^{-\J2\pi(\boldsymbol{\delta} - \delta_1)/\lambda} \Big)^{\ast},  ~ m \in \cM,
\end{align}
where $\sfh_{g,\ell}[m]$ is defined in \eqref{eq:PS_channel_m}.
Finally, given prior knowledge of (i) the distance vector $\boldsymbol{\delta}$ and (ii) the set of codebook angles $\cO$, the RIS interaction codebook can be calculated as
\begin{align}
	\boldsymbol{\cF} & = \left\lbrace \bpsi_m \in \bbC^{N \times 1} : \right. \nonumber \\ & \hspace{8pt} \quad \left. \bpsi_m =   ( \bv (\theta_m) \odot \E^{-\J2\pi(\boldsymbol{\delta} - \delta_1)/\lambda} )^{\ast}, \theta_m  \in \cO \right\rbrace.
\end{align}
where $\left| \boldsymbol{\cF} \right|=\left| \cO \right|=M$.
Given the designed RIS codebook, we formulate next the scene depth map estimation solution.

\begin{algorithm}[h]
	\caption{RIS-Based Scene Depth Estimation Solution}
	\label{alg1}
	\begin{algorithmic}[1]
		
		\Require \begin{varwidth}[t]{\columnwidth} 
			Field of view $\! \rm{FoV}$, aspect ratio $\! A_\rm{R}$, \\ number of horizontal/vertical grid points  $\! \overline{N}_\rm{H},\overline{N}_\rm{V}$.
		\end{varwidth}
		
		\Ensure Depth map estimate $\widehat{\bD}_\rm{map}$.
		
		\State 
		Design RIS interaction codebook $\boldsymbol{\cF}$,  as in \sref{ssec:CB_design}. 		
		
		\For{$m=1$ \textbf{to}  $M$} \!\!\!\!  \Comment{For each $\bpsi_{m}$}	
		\State \begin{varwidth}[t]{\columnwidth} 
			Acquire receive \textit{sensing} signal $z[s,m]$, $\forall s\in \cS$, \eqref{eq:PS_receive_sensing_scalar}.
		\end{varwidth} 
		
		\EndFor
		
		\State Construct receive \textit{sensing} matrix $\bZ$, as in \eqref{eq:sensing_matrix}.
		
		\State 
		Calculate scene range estimate vector $\hat{\br}$, as in \eqref{eq:Scene_range_est}.
		
		\State \begin{varwidth}[t]{\columnwidth} 
			Construct the range map estimate $ \widehat{\bR}_\rm{map}$, as in \eqref{eq:range_map}.
		\end{varwidth}
		
		\State \begin{varwidth}[t]{\columnwidth} 
			Construct the depth map estimate $\widehat{\bD}_\rm{map}$, as in \cite{Taha2021}.	
		\end{varwidth}
		
	\end{algorithmic}
\end{algorithm}

\subsection{Scene Depth Estimation} \label{ssec:scene_range_section}
In this section, we formulate the scene depth map estimation solution, which is outlined in Algorithm \ref{alg1}. 
First, the RIS interaction codebook $\boldsymbol{\cF}$ is designed, as covered in \sref{ssec:CB_design}. 
Then, the sensing system sweeps over the RIS codebook and acquires the receive sensing signal for every RIS interaction vector $\bpsi_m \in \boldsymbol{\cF}$, as defined in \eqref{eq:PS_receive_sensing_scalar}. 
After that, the receive sensing matrix $\bZ$ is constructed as in \eqref{eq:sensing_matrix}.
The receive sensing matrix is then processed using 1D Fourier transforms along its column dimension, to calculate the scene range estimate for every grid point $m \in \cM$. The Fourier-based range profile matrix $\bZ^\rm{RP} \in \bbC^{M_\rm{sample} \times M}$ can be formulated as
\begin{equation}
	\bZ^\rm{RP} = \rm{FFT}_m \left(\bZ \right), m \in \cM, 
\end{equation}  
where $m$ is the column index of the matrix $\bZ$. The scene range estimate vector $\hat{\br} \in \bbR^{M}$ can then be calculated as
\begin{equation}
	\left[ \hat{\br} \right]_{m} = \Delta_\rm{R} \times \argmax_{s} \left|  \left[ \bZ^\rm{RP} \right]_{s,m} \right| , m \in \cM, \label{eq:Scene_range_est}
\end{equation}  
where $\Delta_\rm{R} =  \varsigma/ (2 \rm{BW)}$ is the range resolution.

Next, the sensing system constructs the 2D range map estimate $\widehat{\bR}_\rm{map} \in \bbR^{\overline{N}_\rm{V} \times \overline{N}_\rm{H}}$ from the 1D scene range estimate vector $\hat{\br} \in \bbR^{M}$. Let the horizontal grid index be denoted by $h_\rm{map}\in \{1,\dotsc,\overline{N}_\rm{H}\}$ and the vertical grid index be denoted by $v_\rm{map} \in \{1,\dotsc,\overline{N}_\rm{V}\}$. By converting linear indices to matrix subscripts, the range map estimate $\widehat{\bR}_\rm{map}$ is constructed as
\begin{equation} \label{eq:range_map} 
	\left[ \widehat{\bR}_\rm{map}\right]_{v_\rm{map},h_\rm{map}}  = \left[ \hat{\br} \right]_{m}, \, m=(v_\rm{map}-1)\overline{N}_\rm{H}+h_\rm{map},
\end{equation}
where $m \in \{1,\dotsc,M\}$. After that, the depth map estimate $\widehat{\bD}_\rm{map} \in \bbR^{\overline{N}_\rm{V} \times \overline{N}_\rm{H}}$ can then be calculated from the range map estimate $\widehat{\bR}_\rm{map}$ and the set of angles of the grid points' spherical coordinates, as detailed in \cite{Taha2021}. Finally, the depth map estimate is mapped from the codebook resolution of of $\overline{N}_\rm{V} \times \overline{N}_\rm{H}$ pixels to the desired up-scaled depth map resolution of $M_h \times M_w$ pixels, using 2D signal interpolation \cite{Taha2021}.

\section{Simulation Results}  \label{sec:Sim_Results}

In this section, we evaluate the performance of our proposed RIS based depth map estimation solution.

\subsection{Simulation Framework}

We follow the simulation framework in \cite{Taha2021} to evaluate the performance of the proposed solution with realistic channels.
We first build a detailed floor plan with sufficient number of facets using a high-fidelity 3D graphics design engine, Blender \cite{Blender}. 
this floor plan is then exported to an accurate 3D ray-tracing simulator, Wireless Insite \cite{Remcom}.
Using the ray-tracing output data, we use MATLAB to construct the receive signal models and implement the proposed solution.
For comparison, we generate the ground truth depth map by placing a depth camera in Blender at the same position of the RIS reference element, and adjusting the camera scene parameters to follow the same scene parameters of the RIS codebook grid.

\begin{table}[h]
	\caption{Adopted RIS-aided Sensing System Parameters}
	\begin{center}
		\begin{tabular}{ll}
			\hline
			\textbf{System Configuration}\\
			\hline
			RIS architecture, ${N}_\rm{H}\times {N}_\rm{V}$ &  $\left\lbrace 30 \times 30; 40 \times 40 \right\rbrace $\\
			Starting frequency, $f_{0}$ & $\SI{60}{\giga\hertz}$\\
			Chirp slope, $S$ & $\SI{300}{\mega\hertz \per\micro\second}$\\
			ADC Sampling frequency, $F_{\rm{S}}$ & $\SI{38}{\mega S\per\second}$\\
			Samples per chirp, $M_\rm{sample}$ & $512$\\
			Chirp repetition interval, $T_\rm{PRI}$ & $\SI{13.47}{\micro\second}$\\
			\hline
			\textbf{Derived Parameters}\\
			\hline
			Chirp duration, $T_\rm{active}$ & $\SI{13.47}{\micro\second}$\\
			Transmission bandwidth, $\rm{BW}$ & $\SI{4.04}{\giga\hertz}$\\
			Range resolution, $\Delta_\rm{R}$ & $\SI{3.71}{\centi\meter}$\\
			Maximum range, $\rm{R}_\rm{max}$ & $\SI{18.95}{\meter}$\\
			Chirp rate, $F_\rm{chirp}$ &$\SI{74.2}{\kilo\hertz}$\\
			RIS codebook size, $\left| \boldsymbol{\cF} \right|=M$ &  $\left\lbrace  14,400;\, 25,600 \right\rbrace $\\
			Depth map sensing rate, $F_\rm{DM}$ & $\left\lbrace 5.15, 2.90 \right\rbrace  \SI{}{\hertz}$\\
			\hline
		\end{tabular}
		\label{tb:system_config}
	\end{center}
\end{table}

\noindent \textbf{System Model:}
The RIS-aided sensing system parameters are summarized in Table \ref{tb:system_config}. 
The adopted RIS architectures are $30 \times 30$ and $40 \times 40$ UPAs, i.e. ${N}_\rm{H}={N}_\rm{V}={N}_\rm{RIS} \in \{30,40\}$. For simplicity, assume the radar cross section gain of the RIS elements is an isotropic gain with half-wavelength RIS element spacing, $\sfd=\lambda/2$. The transmit power of the radar system is set to $\SI{20}{\decibel m}$ and $\SI{15}{\decibel m}$  for the $30 \times 30$ and $40 \times 40$ RIS architectures, respectively. The transmit/receive gain of the feeding antenna is assumed to reach a maximum of $\SI{25}{\decibel i}$ in the direction of the RIS elements. The maximum effective isotropic radiated power (EIRP) is then $\SI{45}{\decibel m}$ and $\SI{40}{\decibel m}$ for the adopted RIS architectures, respectively. 

\begin{figure*}[h!]
	\centering
	
	\begin{subfigure}[t]{0.31\textwidth}
		\centering
		\includegraphics[width=\textwidth]{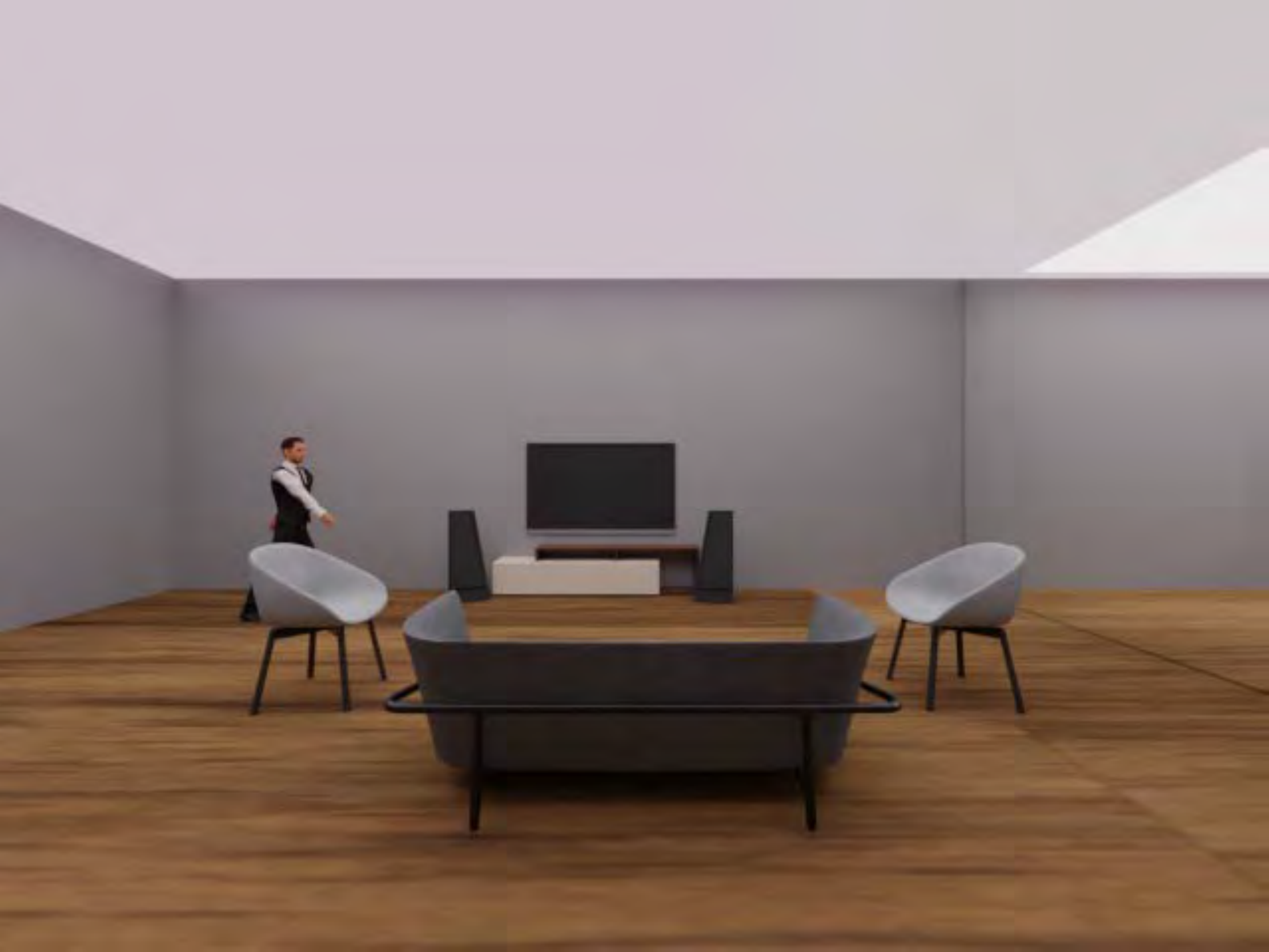}
		\caption{RGB scene image}
		\label{}
	\end{subfigure}
	\hfill 
	\begin{subfigure}[t]{0.31\textwidth}
		\centering
		\includegraphics[width=\textwidth]{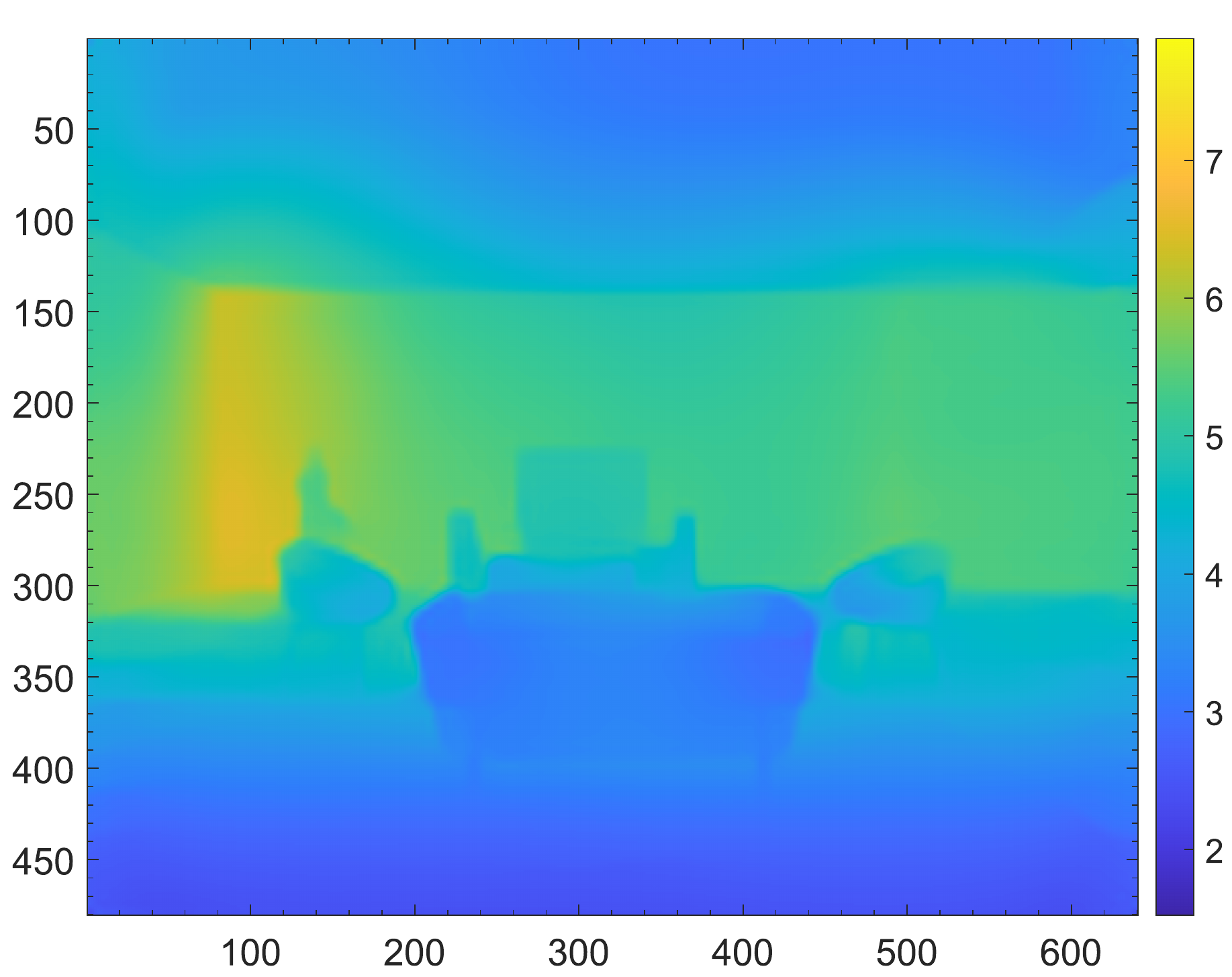}
		\caption{RGB-based depth map \cite{hu2019} \\ $\Delta_{\rm{RMSE}} = \SI{94.5}{\centi\meter}$ \\ $\Delta_{\rm{MAE}} = \SI{82.9}{\centi\meter}$}
		\label{}
	\end{subfigure}
	\hfill 
	\begin{subfigure}[t]{0.31\textwidth}
		\centering
		\includegraphics[width=\textwidth]{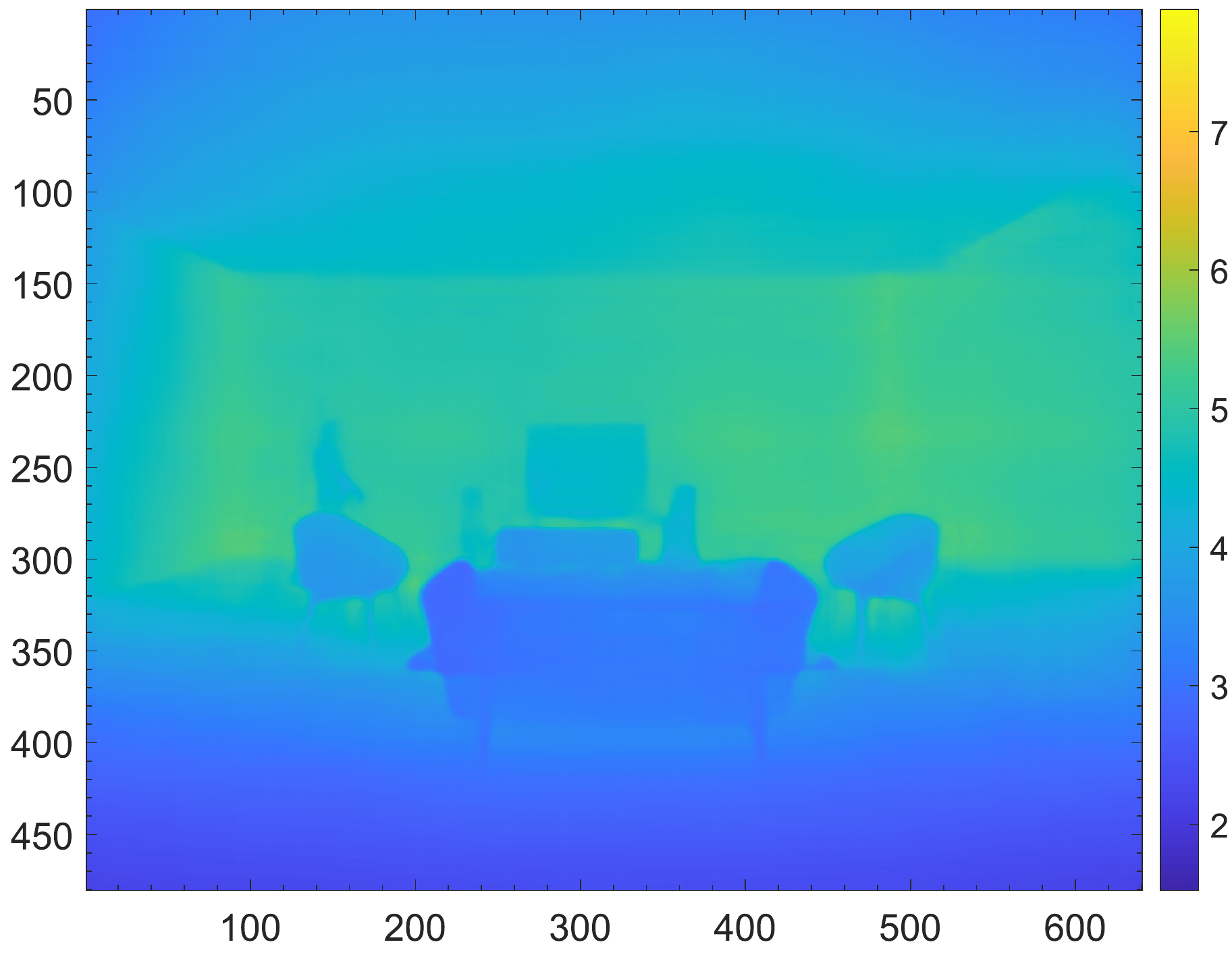}
		\caption{RGB-based depth map \cite{ranftl2021} \\ $\Delta_{\rm{RMSE}} = \SI{98.3}{\centi\meter}$ \\ $\Delta_{\rm{MAE}} = \SI{86.6}{\centi\meter}$}
		\label{}
	\end{subfigure}
	
	\begin{subfigure}[t]{0.31\textwidth}
		\centering
		\includegraphics[width=\textwidth]{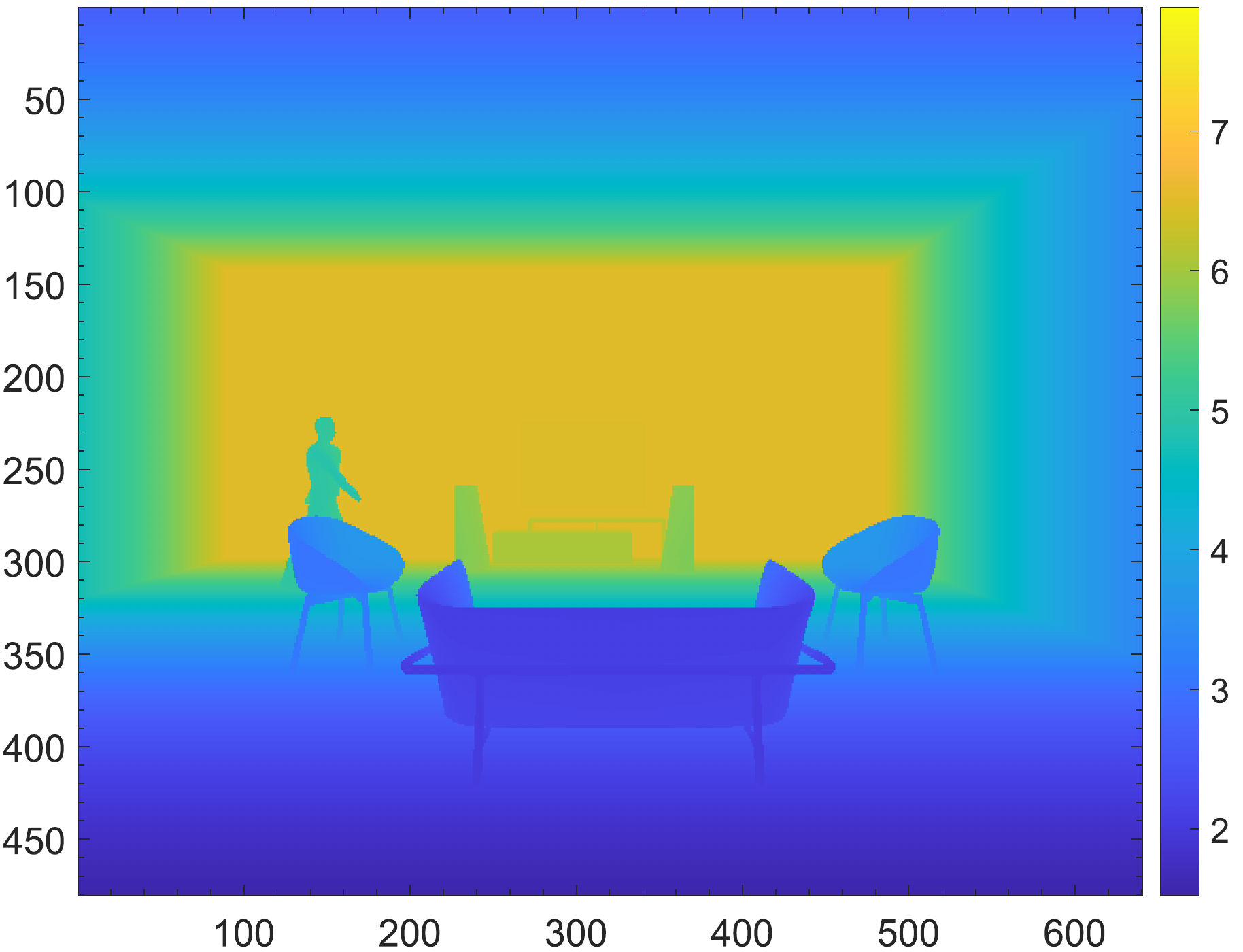}
		\caption{Ground truth depth map}
		\label{}
	\end{subfigure}
	\hfill 
	\begin{subfigure}[t]{0.31\textwidth}
		\centering
		\includegraphics[width=\textwidth]{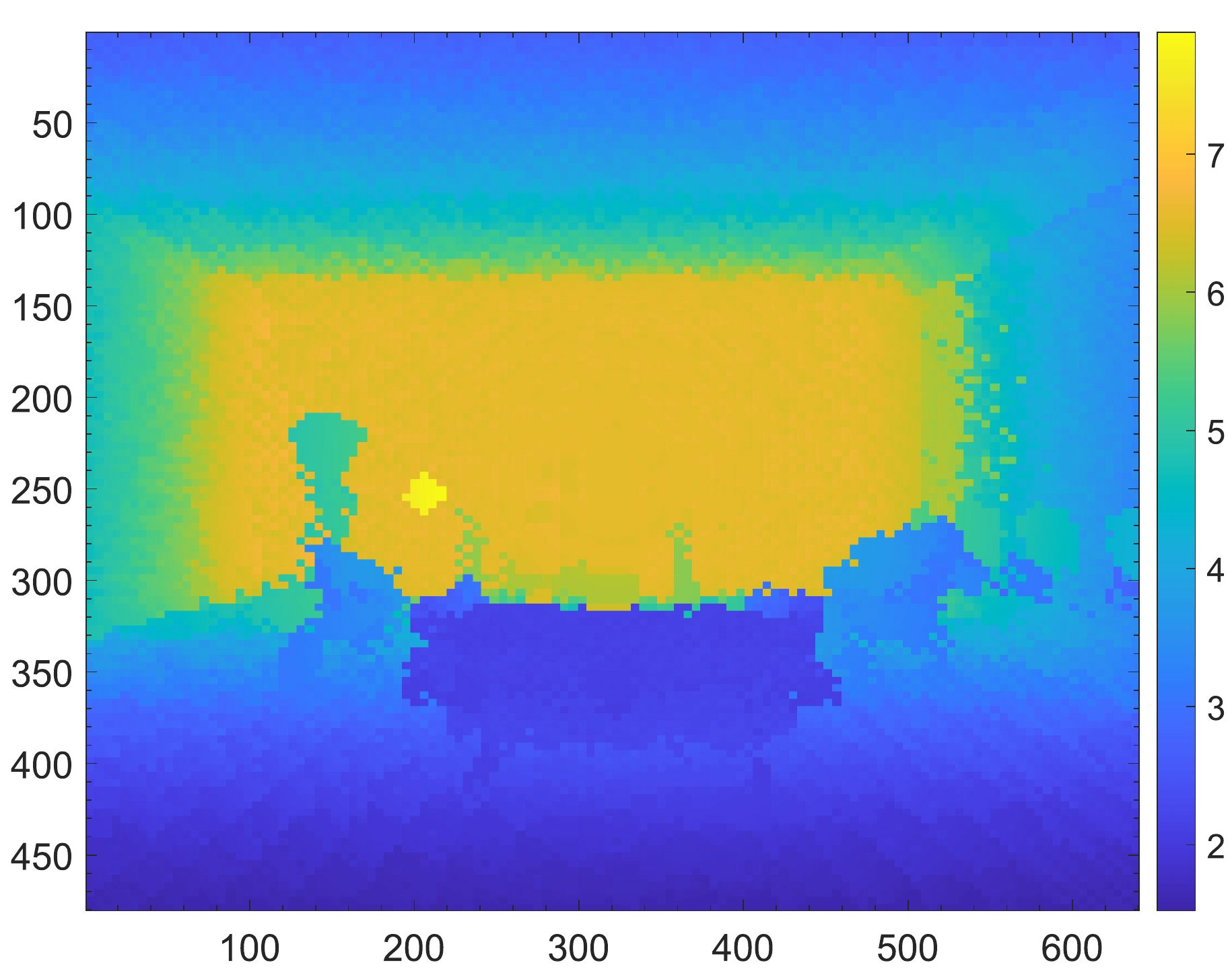}
		\caption{Proposed sol. ($30 \times 30$ RIS) \\ $\Delta_{\rm{RMSE}} = \SI{37.5}{\centi\meter}$ \\ $\Delta_{\rm{MAE}} = \SI{14.5}{\centi\meter}$}
		\label{}
	\end{subfigure}	
	\hfill 
	\begin{subfigure}[t]{0.31\textwidth}
		\centering
		\includegraphics[width=\textwidth]{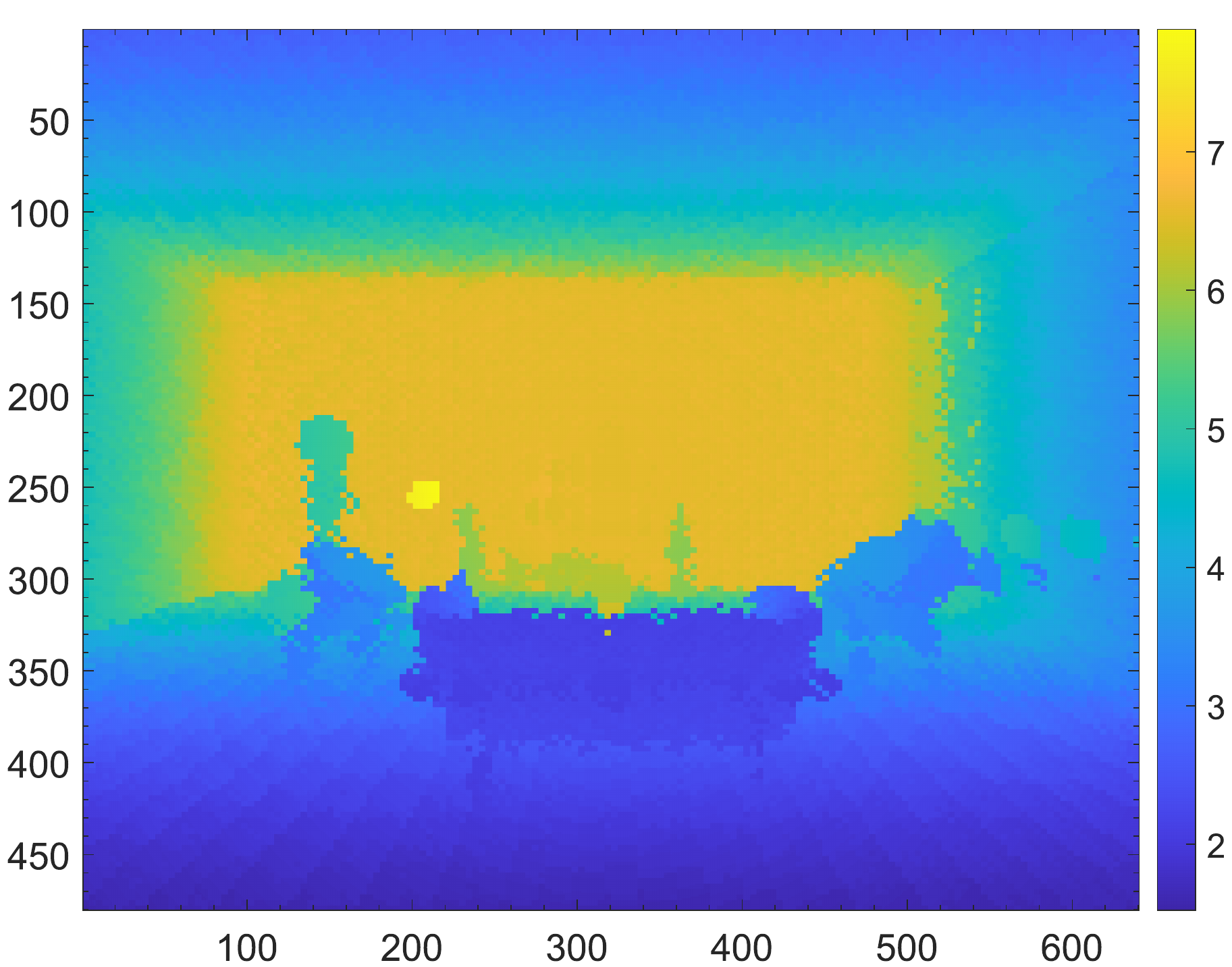}
		\caption{Proposed sol. ($40 \times 40$ RIS) \\  $\Delta_{\rm{RMSE}} = \SI{31.9}{\centi\meter}$ \\ $\Delta_{\rm{MAE}} = \SI{11.6}{\centi\meter}$}
		\label{}
	\end{subfigure}
	
	\caption{For the living room scenario, the proposed RIS-based depth estimation solution is compared against two RGB based depth estimation solutions \cite{hu2019,ranftl2021} and the ground truth depth map. The RIS is equipped with $30 \times 30$ or $40 \times 40$ UPA elements. (a) The scene under study; (b-c) the RGB based estimated maps \cite{hu2019,ranftl2021}; (d) the ground truth map; (e-f) the RIS-based estimated maps.}
	\label{fig:living_room_comparison}
\end{figure*}

\noindent \textbf{Receive Signal Generation:} 
The receive radar signals are generated in two steps.
The first step is generating the parameters of the channel paths using Wireless InSite \cite{Remcom}.
The adopted propagation model and diffuse scattering parameters are the same as the ones in \cite{Taha2021}.
The second step is using the generated channel data to construct the receive IF digital signals \eqref{eq:PS_receive_sensing_scalar}. 
The noise floor is calculated based on the transmission bandwidth and a noise figure of $\SI{10}{\decibel}$. 

\noindent \textbf{RIS-aided Depth Map Estimation Parameters:}
The RIS interaction matrix is designed based on a $100^{\circ}$ field of view centered on the RIS boresight, a $4/3$ scene aspect ratio, and horizontal and vertical oversampling factors of $F_\rm{H}^\rm{OS} = F_\rm{V}^\rm{OS} = 4$. The RIS codebook size is calculated using $\left| \boldsymbol{\cF} \right|=M= {N}_\rm{H} F_\rm{H}^\rm{OS} {N}_\rm{V} F_\rm{V}^\rm{OS}$. Correspondingly, the ground truth depth maps are generated by Blender with a $100^{\circ}$ field of view and a sensor width of $\SI{32}{\milli\meter}$. The image quality of the ground truth depth maps and the up-scaled estimated depth maps is set to $480\rm{p}$ resolution, i.e. $640\times480$ pixels. Next, we evaluate the performance of our proposed RIS aided depth map estimation solution in an indoor living room scenario.

\subsection{Results for A Living Room Scenario}

In this scenario, we consider a $\SI{15.6}{\meter} \times \SI{6.5}{\meter} \times \SI{3.8}{\meter}$ indoor space, with a glass wall dividing the space into two rooms.
The room under study is a $\SI{9.6}{\meter} \times \SI{6.5}{\meter} \times \SI{3.8}{\meter}$ living room, where a $\SI{1.8}{\meter}$ tall person is moving from left to right. 
The adopted materials of the inanimate objects/surfaces follow the ITU default parameter values at $60$GHz and are classified into concrete, floorboard, ceiling board, glass, or wood.
The RIS is assumed to be placed on the wall behind the sofa. 
The number of facets ranges between $\approx 2\text{k}$ and $30\text{k}$ for the inanimate objects/surfaces. $20,542$ facets are used for the person model.
We compare the proposed solution against two RGB-based solutions \cite{hu2019,ranftl2021} to demonstrate the capability of the RIS-aided sensing in 
\begin{enumerate*}[label={(\roman*)}]
	\item detecting transparent surfaces and
	\item achieving higher trueness of the estimated depth values.
\end{enumerate*}
We follow the official implementation of these RGB-based solutions and utilize the well-trained models on the NYU depth V2 dataset \cite{silberman2012}.

\figref{fig:living_room_comparison} compares the estimated depth maps from the RIS-based solution against the ones from the RGB-based solutions, which uses monocular RGB images to estimate the depth maps.
As shown, the RGB-based solutions can construct the shape of the objects/person more clearly than the proposed solution; i.e they achieve a higher depth precision.
These RGB-based solutions, however, do not achieve high depth accuracy due to their low level of trueness, especially when misdetecting the transparent glass wall.
As for the proposed RIS-based solution, even though the glass has the lowest scattering factor among all the other materials, the depth of the glass wall can be better perceived with a lower estimation error; i.e. the proposed solution can achieve a higher depth trueness.
the depth estimation accuracy of the RIS-based solution, however, suffers from inter-path interferences at some directions, where the receive powers of undesired paths are higher than the ones of the desired single-bounce paths. 
Although the RIS based solution offers a higher spatial resolution than the mmWave MIMO based solution \cite{Taha2021}, the RIS reflected beams are yet relatively wide compared to the ideal pencil beams. For this reason, high estimation errors are observed around the edges of the objects/person. 
It would be interesting to address these challenge in future work. 

\section{Conclusion}  \label{sec:Conclusion}
In this paper, we considered the problem of scene depth map estimation using mmWave wireless sensing systems. 
For this problem, we proposed to leverage RISs to accurately estimate high-resolution depth maps. 
To achieve this objective, we formulated the RIS wireless sensing based scene depth estimation problem and proposed a comprehensive framework for building scene depth maps using RIS aided mmWave sensing systems. 
The proposed framework includes designing an RIS interaction codebook capable of creating a sensing grid of reflected beams that meets the desirable characteristics of efficient scene depth map construction. 
Using the designed RIS interaction codebook, a post-processing solution is developed to build high-resolution depth maps.	
Adopting accurate 3D ray-tracing models, the results showed that the developed solution can achieve depth map estimation errors in the order of $\SI{12}{\centi\meter}$. 
This highlights the potential of leveraging this proposed solution in achieving accurate depth perception of the surrounding environment.

\bibliographystyle{IEEEtran}

\end{document}